\documentclass[lettersize,journal]{IEEEtran}
\usepackage{amsmath,amsfonts}
\usepackage{algorithmic}
\usepackage{algorithm}
\usepackage{array}
\usepackage[caption=false,font=normalsize,labelfont=sf,textfont=sf]{subfig}
\usepackage{textcomp}
\usepackage{stfloats}
\usepackage{url}
\usepackage{verbatim}
\usepackage{graphicx}
\usepackage{cite}
\usepackage{cuted}
\usepackage{lipsum}
\begin{document}

\title{A Precoding for ORIS-Assisted MIMO Multi-User VLC System}

\author{Mahmoud Atashbar, Hamed Alizadeh Ghazijahani, Yong Liang Guan,~\IEEEmembership{Senior Member, IEEE}, and Zhaojie Yang
\thanks{This paper was produced by the IEEE Publication Technology Group. They are in Piscataway, NJ.}
\thanks{}}


\IEEEpubid{IEEE}

\maketitle

\begin{abstract}
In this paper, we study a multi-user visible light communication (VLC) system assisted with optical reflecting intelligent surface (ORIS). Joint precoding and alignment matrices are designed to maximize the average signal-to-interference plus noise ratio (SINR) criteria. Considering the constraints of the constant mean transmission power of LEDs and the power associated with all users, an optimization problem is proposed. To solve this problem, we utilize an alternating optimization algorithm to optimize the precoding and alignment matrices. The simulation results demonstrate that the resultant SINR of the proposed method outperforms ZF and MMSE precoding algorithms.
\end{abstract}

\begin{IEEEkeywords}
VLC, Precoding, ORIS, Multi-user, MIMO.
\end{IEEEkeywords}

\section{Introduction}
\IEEEPARstart{C}{ommunicating} data using visible light is an attractive optical communication system that utilizes the visible region of the optical spectrum. As an emerging technology in 6G, visible light communication (VLC) combines communication with lighting and is a promising technique for providing indoor internet access. VLC offers several significant advantages, such as license-free spectrum, high security, high data rates, low cost, and no hazardous electromagnetic radiation\cite{wang2017visible}. 

Multiple-input multiple-output (MIMO) technology is becoming popular in VLC systems. In a multi-user MIMO VLC (MU-MIMO-VLC) system, the transmitter is an array of transmitting LEDs that simultaneously supports multiple users  equipped with single or multiple photodiodes (PD). To mitigate inter-user interference, the signals associated with the users undergo precoding prior to transmission\cite{wang2022joint}.
The main objective of precoding is to maximize the signal-to-interference plus noise ratio (SINR) at the receivers. By directing the transmitted signals toward the intended users, it becomes feasible to alleviate the impacts of multi-user interference in MU-MIMO-VLC systems such as zero-forcing (ZF) and minimum mean square error (MMSE) precoding algorithms.

On the other hand, reflecting intelligent surface (RIS) technology opens a wide insight into wireless communications that enables cost-energy and spectral-efficient communications \cite{mohammadi2023comprehensive}. Often, in radio frequency communications (RF), these surfaces incorporate phase shifters to reflect electromagnetic waves with specific phases according to specific inputs or requirements. To get the benefits of deploying RISs in wireless networks, the RIS parameters such as phase shift values and network resource allocation like precoding, must be optimized jointly \cite{rf_ris}. In the VLC system, the optical RISs (ORIS) are utilized that reflect the incident signals intelligently \cite{road_tvt}.
The ORIS can compensate for path loss and improve the overall signal strength at the receivers. From the physics standpoint, the principal classifications of ORISs comprise the intelligent meta-surface reflector, intelligent mirror array, and liquid crystal-based ORIS \cite{RIS-VLC_survey}.

Some works address MIMO-VLC in the presence of ORIS. The authors in \cite{sun2023intelligent} designed jointly the precoding and detection matrices for a multi-PD single user by minimizing the receiver MMSE. In \cite{mushfique2022mirrorvlc} the transmit power of LEDs and the location of mirrors are designed based on maximizing SINR for a multi-user scenario. The alignment matrix between LEDs and ORIS elements and transmit power are designed by maximizing the achievable rate for multi-user scenario in \cite{sun2023capacity}. Another work presents a codebook-assistant precoding for an omni-digital ORIS with single user case \cite{ndjiongue2023double}.
 
In this paper, we design precoding and alignment matrices jointly for the ORIS-assisted MU-MIMO-VLC system that maximizes the mean SINR of users under the following constraints. Each element in ORIS reflects the light to only one user, so the mean alternative current (AC) power of LEDs and the allocated power of all users are the same. To solve the proposed optimization problem, we propose an alternating algorithm. By supposing a known alignment matrix, the gradient ascent method that is projected on Manifolds is utilized to design the precoding matrix. Considering a known precoding matrix, the alignment matrix is designed using a low-complexity method.

\section{System Model}
\begin{figure}[!t]
\centering
\includegraphics[width=0.75\linewidth]{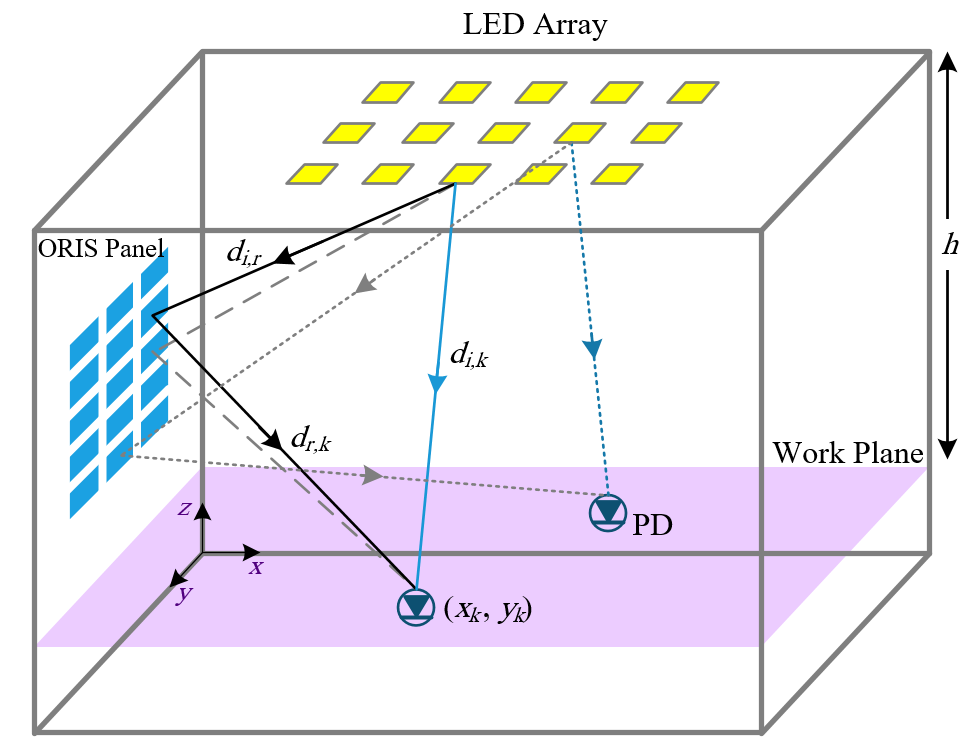}
\caption{The MIMO-VLC system model with an ORIS installed on a wall.}
\label{system_model}
\end{figure}

In this work, we consider a room with a rectangular LED array on the ceiling and a rectangular ORIS panel with $M$ elements installed on the wall to model a MU-MIMO VLC system, shown in Fig. \ref{system_model}. The LED array with $N_t$ LEDs supports $K$ single-PD users placed on the work plane simultaneously. Each user receives the optical signal from the line-of-sight (LOS) channel emitted from the LED panel and the signal is reflected from the ORIS panel, as a non-LOS (nLOS) channel. The LOS channel between $i$th LED and $k$th user follows the well-known Lambertian model as \cite{wang2017visible} 
\begin{equation} 
\label{lambertian1}
h^{(los)}_{i, k}=\frac{A_p(m+1)}{2 \pi d_{i, k}^2} \cos ^m(\phi_{i,k})\cos (\theta_{i,k}) TG,
\end{equation} 
in which $\phi_{i,k}$ is the emitting angle, $\theta_{i,k}$ denotes the incident angle from the $i$th LED to $k$th user, $m$ is Lambert’s mode number expressing directivity of the source beam, $A_p$ denotes the area of the receiver PD, the $T$ represents the signal transmission coefficient of an optical filter, $G$ denotes the concentrator gain, and $d_{i,k}$ is the distance between $i$th LED to $k$th user \cite{wang2017visible}. Note that (\ref{lambertian1}) is valid for $0<\theta_{i,k}<\theta_0$ with $\theta_R$ as semi-angle of the PD field-of-view.

The nLOS channel for the signal transmitted from $i$th LED, reflected  from $r$th element of ORIS, and received with $k$th user is indicated as \cite{abdelhady2020visible}
\begin{equation} 
\label{lambertian2}
h^{(nlos)}_{i,r, k}=\alpha\frac{A_p(m+1)}{2 \pi (d_{i, r}+d_{r,k})^2} \cos ^m(\phi_{i,r})\cos (\theta_{r,k}) TG,
\end{equation} 
where $0<\alpha<1$ is the reflective index of the ORIS.

Supposing $\boldsymbol{s}=\left[s_1, s_2, \ldots, s_K\right]^T \in R^{K \times 1}
$ is the transmitted signal vector where $s_i$ indicates the transmitted symbol to $k$th user. At first, the transmitted signal vector is multiplied by the precoding matrix $
\boldsymbol{P}\epsilon R^{N_t \times K}$ with $\boldsymbol{P}=\left[\boldsymbol{p}_1, \boldsymbol{p}_2, \ldots, \boldsymbol{p}_K\right]$
where $\boldsymbol{p}_k$ is the precoding vector associated with $k$th user to generate the LEDs' transmission signal vector. Then, the DC offset is summed up with the precoded signal to guarantee the non-negative transmission signal of LED's.

On the receiver side, the received signal of the $k$th user after removing  the DC value is
\begin{equation}
\label{rec. signal. model}
y_k=\left(\boldsymbol{h}_k^{(\text {los })}+\boldsymbol{H}_k^{(\mathrm{nlos})} \boldsymbol{f}_k\right)^{\boldsymbol{T}} \boldsymbol{P} \boldsymbol{s}+n_k,
\end{equation}
where $n_k$ is the zero-mean Gaussian noise parameter with variance $\sigma_n^2$ and
$\boldsymbol{h}_k^{(\mathrm{los})}=\left[h_{1, k}^{(\mathrm{los})}, h_{2, k}^{(\mathrm{los})}, \ldots h_{N_t, k}^{(\mathrm{los})}\right]^T$
and  $\boldsymbol{H}_k^{(\mathrm{nlos})} \boldsymbol{f}_k$ are LOS and nLOS components of the channel vector, respectively where $\boldsymbol{H}_k^{(\mathrm{nlos})} $ is the channel matrix associated with LED and ORIS arrays for $k$th user as
$\boldsymbol{H}_k^{(\text {nlos })}=\left[\boldsymbol{h}_{1, k}^{(\text {nlos) }}, \boldsymbol{h}_{2, k}^{\text {(nlos) }}, \ldots, \boldsymbol{h}_{N_t, k}^{\text {(nlos) }}\right]^T$,
with $\boldsymbol{h}_{i, k}^{(\text {nlos })}=\left[h_{i, 1, k}^{\text {(nlos) }}, h_{i, 2, k}^{(\text {nlos })}, \ldots h_{i, M, k}^{(\text {nlos })}\right]^T$.
It is supposed that any element in ORIS is aligned with a user and correspondingly each user receives the nLOS component of signals via determined elements of ORIS that are aligned with. Define $\boldsymbol{f}_k=\left[f_{1, k}, f_{2, k}, \ldots, f_{M, k}\right]^T$ as an alignment binary vector associated with $k$th user, where $f_{r, k} \in\{0,1\}$ for $r=1, 2, ..., M$. If element $r$ of ORIS is aligned with $k$th user, $f_{r, k}=1$, otherwise $f_{r, k}=0$. Considering $M_k$ as the number of ORIS elements aligned with $k$th user, we have $\sum_{r=1}^M f_{r, k}=M_k$ besides $\sum_{k=1}^K M_{k}=M$.

\section{Proposed Precoding Method}
\subsection{proposed optimization problem}
We propose an optimization problem to joint design the precoding matrix $\mathbf{P}$ and alignment matrix 
$\boldsymbol{F}=[\boldsymbol{f_1}, \boldsymbol{f_2}, ..., \boldsymbol{f_K}]$.
To this end, we reformulate the (\ref{rec. signal. model}) as
\begin{align}
\begin{split}
\label{res-int}
y_k=&\underbrace{\left(\boldsymbol{h}_k^{\text {(los) }}+\boldsymbol{H}_k^{(\text {nlos) }} \boldsymbol{f}_k\right)^{\boldsymbol{T}} \boldsymbol{p}_k s_k}_{\text {main term }}\\
&+\underbrace{\sum_{\substack{j=1 \\ j \neq k}}^K\left(\boldsymbol{h}_k^{(\text {los) }}+\boldsymbol{H}_k^{(\text {nlos) }} \boldsymbol{f}_k\right)^T \boldsymbol{p}_j s_j}_{\text {interference term }}+n_k.
\end{split}
\end{align}

In (\ref{res-int}), the first term is the target signal we aim to receive at user $k$ and the second one is the interference signal associated with other users. Assuming that users' signals are independent of each other and noise, the  $k$th user SINR can be written as
\begin{equation}
\label{sinr}
\text{SINR}_{k}=\frac{\left\|\left(\boldsymbol{h}_k^{(\mathrm{los})}+\boldsymbol{H}_k^{(\mathrm{nlos})} \boldsymbol{f}_k\right)^T \boldsymbol{p}_k\right\|^2}{\sum_{\substack{j=1 \\ j \neq k}}^K\left\|\left(\boldsymbol{h}_k^{(\mathrm{los})}+\boldsymbol{H}_k^{(\mathrm{nlos})} \boldsymbol{f}_k\right)^T \boldsymbol{p}_j\right\|^2+\boldsymbol{\sigma}_n^2}.
\end{equation}

In the proposed optimization problem, the idea is to design a precoding and an alignment matrix to maximize the $\text{SINR}$ under the constraints of the alignment matrix. The overall $\text{SINR}$ is defined as the mean  $\text{SINR}$ value of users as
\begin{equation}
\label{sinr_mean}
    \text{SINR} = \frac{1}{K}\sum_{k=1}^{K} \text{SINR}_k.
\end{equation}

To achieve a higher power efficiency of the amplifiers used to drive the LED array, it is needed that the mean transmitted AC powers of the signals associated with all LEDs be the same. According to the unit mean power assumption of vector $\boldsymbol{s}$ elements, the declared constraint becomes 
$ \sum_{k=1}^{K}{p}_{i,k}=1, \forall{i=1,2, ... N_t}$
 where $p_{i,k}$ represents the $i$th element of $\boldsymbol{p}_k$. This constraint can be expressed in the matrix form as $\text {diag} \left(\boldsymbol{P} \boldsymbol{P}^{\text {H }} \right) =\boldsymbol{I}_{N_t}$, where $\boldsymbol{I}_{N_t}$ is $N_t \times N_t $ identity matrix. Moreover, the constant associated power with all users is considered as another constraint that leads to 
 $ \sum_{i=1}^{N_t}{p}_{i,k}$ 
 being constant for all users. Simply it can be shown that the referred constant value is $\frac{N_t}{K}$. The resultant constraint in matrix form is
 $\text {diag} \left(\boldsymbol{P}^{\text {H }}\boldsymbol{P} \right)=\frac{N_t}{K}\boldsymbol{I}_{K}$.
 Consequently, the proposed optimization problem is summarized as
\begin{subequations}
\begin{alignat}{1}
\text{(P1):} &\max_{\boldsymbol{P,F}} \text{ SINR,}\\
\text{s.t. \          }&f_{k m} \epsilon\{0,1\},\label{cons1}\\
&\sum_{m=1}^M f_{k m}=M_k,\label{cons2}\\
&\sum_{k=1}^K M_k=M,\label{cons3}\\
&\text {diag} \left(\boldsymbol{P} \boldsymbol{P}^{\text {H }} \right)=\boldsymbol{I}_{N_t},\label{cons4}\\
&\text {diag} \left(\boldsymbol{P}^{\text {H }}\boldsymbol{P} \right)=\frac{N_t}{K}\boldsymbol{I}_{K}.\label{cons5}
\end{alignat}
\end{subequations}
    
Subsequently, we will propose a method to solve (P1).

\subsection{solving the optimization problem}
Note that (P1) is an NP-hard problem. Solving this optimization problem to find precoding and alignment matrices jointly is very difficult. Therefore, we propose an alternating optimization algorithm in this paper to solve the problem so $\boldsymbol{F}$ and $\boldsymbol{P}$ are iteratively optimized. In the optimization algorithm, assuming that the value of one of the matrices is fixed, we optimize the other one, and in the next step, vice versa. We repeat this procedure until the algorithm converges. Supposing an initialization random precluding matrix, the general algorithm for joint precoding and alignment matrices design is as algorithm \ref{general_algorithm}.

\begin{algorithm}[H]
\caption{General optimization algorithm.}\label{general_algorithm}
\begin{algorithmic}
\STATE {\textbf{Input:} $\boldsymbol{h}_k^{\text {(los)}}, \boldsymbol{H}_k^{\text {(nlos)}}\forall{ k=1,2, ... K}$}
\STATE {\textbf{Initialize: }$\boldsymbol{P}\epsilon R^{N_t \times K}$ }
\STATE {\textbf{repeat}}
\STATE \hspace{0.4cm}{run the alignment matrix design algorithm (Algorithm \ref{alg_f})}
\STATE \hspace{0.4cm}{run the precoding matrix design algorithm (Algorithm \ref{precoding_algorithm})}
\STATE  {\textbf{until }convergence}
\STATE{\textbf{Output:} $\boldsymbol{P}$, $\boldsymbol{f}_k \forall{ k=1,2, ... K}$}
\end{algorithmic}
\end{algorithm}

\subsubsection{Alignment matrix design}
Considering the precoding matrix $\boldsymbol{P}$ is given, the optimization problem (P1) is simplified into the following sub-problem 
\begin{equation}
   \text{(P2):           } \max_{\boldsymbol{F}} \text{ SINR, }
\end{equation}
$$\text{s.t.  (\ref{cons1}), (\ref{cons2}) and (\ref{cons3}).}$$

According to the constraint of the problem (P2), $\boldsymbol{F}$ is a binary matrix, so each row contains only a single element with the value of "1", and column $k$ consists of $M_k$ elements with the value of "1". The classical method to find such a binary matrix is to try all possible matrices and choose the best one that maximizes the $\text{SINR}$. Such a solution suffers from computational complexity due to the huge number of possible states. The number of all possible alignment matrices according to the constraints  (\ref{cons1}), (\ref{cons2}), and (\ref{cons3}) is $K^M$ which leads to $K^{M+1}$ times calculating (\ref{sinr}). For instance, let the number of users be $K=\text{4}$ and there are $M=\text{36}$ ORIS elements, so the $\text{SINR}$ must be checked 4.7e+21 times. We propose a method to solve (P2) to cope with this complexity. 

In the proposed method, $r$th element of ORIS is associated with $k$th user so that the $\text{SINR}_k$ related to this element given in (\ref{sinr-rk}) is maximum.  
\newcounter{mytempeqncnt}
\begin{figure*}[!b]
\normalsize 
\setcounter{mytempeqncnt}{\value{equation}}
\setcounter{equation}{8}
\begin{equation}\label{sinr-rk}
\text{SINR}_{k,r}=\frac{\boldsymbol{h}_k^{(\mathrm{los}) T} \boldsymbol{p}_k \boldsymbol{p}_k^{\mathrm{T}} \boldsymbol{h}_k^{(\mathrm{los})}+\mathbf{2}\left[\boldsymbol{H}_k^{(\mathrm{nlos}) T} \boldsymbol{p}_k \boldsymbol{p}_k^{\mathrm{T}} \boldsymbol{h}_k^{(\mathrm{los})}\right]_{\boldsymbol{r}}+\left[\boldsymbol{H}_k^{(\mathrm{nlos}) T} \boldsymbol{p}_k \boldsymbol{p}_k^{\mathrm{T}} \boldsymbol{H}_k^{(\mathrm{nlos})}\right]_{r, \boldsymbol{r}}}{\sum_{\substack{j=1 \\ j \neq k}}^K\left(\boldsymbol{h}_k^{(\mathrm{los}) T} \boldsymbol{p}_j \boldsymbol{p}_j^{\mathrm{T}} \boldsymbol{h}_k^{(\mathrm{los})}+\mathbf{2}\left[\boldsymbol{H}_k^{(\mathrm{nlos}) T} \boldsymbol{p}_i \boldsymbol{p}_i^{\mathrm{T}} \boldsymbol{h}_k^{(\mathrm{los})}\right]_r+\left[\boldsymbol{H}_k^{(\mathrm{nlos}) T} \boldsymbol{p}_i \boldsymbol{p}_i^{\mathrm{T}} \boldsymbol{H}_k^{(\mathrm{nlos})}\right]_{\boldsymbol{r}, \boldsymbol{r}}\right)+\boldsymbol{\sigma}_{\boldsymbol{n}}^2}.    
\end{equation}
\end{figure*}
Considering $\mathcal{R}$ as the set of unassigned ORIS elements indices, we have
\begin{equation}\label{r-star}
r^*=\underset{r \in \mathcal{R}}{\operatorname{argmax}}\text{ SINR}_{k,r}.
\end{equation}

Afterward, we set the $r^*$th element of $\boldsymbol{f}_k$ to one. Consequently, the proposed algorithm to design the alignment matrix is summarized in Algorithm \ref{alg_f}. Using this algorithm, it is needed $\frac{M(M+1)}{2}$ times calculating (\ref{sinr-rk}).

\begin{algorithm}[H]
\caption{Design the Alignment Matrix.}\label{alg_f}
\begin{algorithmic}
\STATE {\textbf{Input:} $\boldsymbol{P}, \boldsymbol{h}_k^{\text {(los)}}, \boldsymbol{H}_k^{\text {(nlos)}}\forall{ k=1,2, ... K}$}
\STATE {\textbf{Initialize: }$\mathcal{R}=\{1,2, ... M\}, \boldsymbol{f}_k=\boldsymbol{0}_{M\times1}\forall{ k=1,2, ... K}$}
\STATE {\textbf{for} $r=1:M$ \textbf{do}}
\STATE \hspace{0.5cm} {\textbf{for} $k=1:K$ \textbf{do}}
\STATE \hspace{1cm}{solve (\ref{r-star}) to find $r^*$}
\STATE \hspace{1cm}{set the $r^*$th element of $\boldsymbol{f}_k$ to one}
\STATE \hspace{1cm}{$\mathcal{R}=\mathcal{R}-\{r^*\}$}
\STATE \hspace{0.5cm} {\textbf{end for}}
\STATE  {\textbf{end for}}
\STATE{\textbf{Output:} $\boldsymbol{f}_k \forall{ k=1,2, ... K}$}
\end{algorithmic}
\end{algorithm}

\subsubsection{Precoding matrix design}

Considering the alignment matrix $\boldsymbol{F}$ is given, the optimization problem (P1) is simplified into the following sub-problem 
\begin{equation}
   \text{(P3):           } \max_{\boldsymbol{P}} \text{ SINR,}
\end{equation}
$$\text{ s.t. (\ref{cons4}) and (\ref{cons5}).}$$

In (P3), SINR is the sum of fractional programming (FP)  in which each is a non-convex function. To deal with this difficulty, we use quadratic transform \cite{fractional} to solve the problem (P3).
As our problem is the sum of $K$ multiple-ratio terms, the idea of decoupled optimization of numerators and denominators to the sum-of-ratio problem must be applied and use a straightforward extension of Dinkelbach’s transform which leads to the following optimization problem
\begin{equation}\label{p4}
\text{(P4): } \max _{\boldsymbol{P}, \boldsymbol{\gamma}} g(\boldsymbol{P}, \boldsymbol{\gamma}),
\end{equation}
$$ \text{  s.t.  (\ref{cons4}) and (\ref{cons5}}),$$
in which
\begin{equation}\label{g_function}
    g(\boldsymbol{P}, \boldsymbol{\gamma})=\sum_{k=1}^K(2 \gamma_k\left|\boldsymbol{h}_k^T \boldsymbol{p}_k\right|-\gamma_{{k}}^2 \sum_{\substack{j=1 \\ j \neq k}}^K\left\|\boldsymbol{h}_k^T \boldsymbol{p}_j\right\|^2-{\gamma^2_k} {\sigma^2_n})
\end{equation}
and $\boldsymbol{\gamma}=\left[\gamma_1, \gamma_2, \ldots  \gamma_K\right]^T$ is a temporary vector that is updated iteratively.

Recently, geometric solutions have been used to solve various optimization problems. One type of solution that is commonly employed in constrained optimization problems is manifold-based geometry \cite{manifold_2008}. This approach is favored due to its relative simplicity and optimality. The constraints in constrained optimization problems can be interpreted as isolated points in the space in manifold forms such as Stiefel, Grassmann, Riemannian, etc \cite{manifold_2008, manton2020geometry}. Accordingly, the optimum points are searched in the space inside the manifold. 
As both the constraints in problem (P4) are in the form of a Grassmann manifold, we propose a gradient ascent method projected on the manifold to solve the problem. In this way, $\boldsymbol{P}$ and $\boldsymbol{\gamma}$ are calculated iteratively as
\begin{subequations}
   \begin{equation}\label{iter-p}
    \boldsymbol{P}_{\text{new}}= \boldsymbol{P}_{\text{old}}+\mu\nabla_{_{\boldsymbol{P}}} g(\boldsymbol{P}, \boldsymbol{\gamma}),
\end{equation} 
\begin{equation}\label{iter-gamma}
    \boldsymbol{\gamma}_{\text{new}}= \boldsymbol{\gamma}_{\text{old}}
    +\mu\nabla_{\boldsymbol{\gamma}} g(\boldsymbol{P}, \boldsymbol{\gamma}),
\end{equation}
\end{subequations}
where $\nabla_{_{\boldsymbol{P}}}$ and $\nabla_{_{\boldsymbol{\gamma}}}$ represent the gradients of a function with respect to $\boldsymbol{P}$ and $\boldsymbol{\gamma}$, respectively and $\mu$ is the step size. Accordingly, it can be shown that the $k$th column of $\nabla_{_{\boldsymbol{P}}} g(\boldsymbol{P}, \boldsymbol{\gamma})$ and $k$th element of vector $\nabla_{\boldsymbol{\gamma}} g(\boldsymbol{P}, \boldsymbol{\gamma})$ are
\begin{subequations}
\begin{equation}\label{grad-p}
\left[\nabla_{_{\boldsymbol{P}}} g(\boldsymbol{P}, \boldsymbol{\gamma})\right]_k=\frac{\gamma_k}{\left|\boldsymbol{h}_k^T \boldsymbol{p}_k\right|} \boldsymbol{h}_k\boldsymbol{h}_k^T \boldsymbol{p}_k-2 \sum_{\substack{j=1 \\ j \neq k}}^K \gamma_{i}^2\boldsymbol{h}_i \boldsymbol{h}_i^T \boldsymbol{p}_k,
\end{equation}
\begin{equation}\label{grad-gamma}
    \nabla_{\gamma_k} g(\boldsymbol{P}, \boldsymbol{\gamma})=2\left|\boldsymbol{h}_k^T \boldsymbol{p}_k\right|-2\gamma_{k} \sum_{\substack{j=1 \\ j \neq k}}^K\left\|\boldsymbol{h}_k^T \boldsymbol{p}_j\right\|^2-2{\gamma_k}\sigma^2_n.
\end{equation}
\end{subequations}

To satisfy $\boldsymbol{P}$ on Grassman manifolds of constraints (\ref{cons4}) and (\ref{cons5}), in each iteration of the gradient ascent algorithm $\boldsymbol{P}_\text{new}$ is projected on the intersection of two manifolds. The projections onto the manifold (\ref{cons4}) and (\ref{cons5}) are as below, respectively.
\begin{equation}\label{proj_G1}
\mathcal{P}_{\mathbb{G} 1}(\boldsymbol{P})=\left(\operatorname{diag}\left(\boldsymbol{P} \boldsymbol{P}^H\right)\right)^{-\frac{1}{2}} \boldsymbol{P},
\end{equation}
\begin{equation}\label{proj_G2}
\mathcal{P}_{\mathbb{G} 2}(\boldsymbol{P})=\sqrt{\frac{N_t}{K}}\boldsymbol{P}\left(\operatorname{diag}\left(\boldsymbol{P}^H \boldsymbol{P}\right)\right)^{-\frac{1}{2}}.
\end{equation}

Projection onto the intersection of two manifolds is given by\cite{lu2020omnidirectional}
\begin{equation}\label{inter_sec}
    \mathcal{P}_{\mathbb{G} 1 \cap \mathbb{G} 2}(\boldsymbol{P})=\mathcal{P}_{\mathbb{G} 1}(\mathcal{P}_{\mathbb{G} 2}(... \mathcal{P}_{\mathbb{G} 1}(\mathcal{P}_{\mathbb{G} 2}(\boldsymbol{P})))).
\end{equation}

Equation (\ref{inter_sec}) shows that the projection of manifolds proceeds alternatively until convergence is reached. Generally, in practice convergence is satisfied in the first five iterations. Algorithm \ref{precoding_algorithm} summarizes the design of the matrix $\boldsymbol{P}$ procedure.
\begin{algorithm}[H]
\caption{Design the Precoding Matrix.}\label{precoding_algorithm}
\begin{algorithmic}
\STATE {\textbf{Input:} $\boldsymbol{f_k}, \boldsymbol{h}_k^{\text {(los)}}, \boldsymbol{H}_k^{\text {(nlos)}}\forall{ k=1,2, ... K}$}
\STATE {\textbf{Initialize: }$\boldsymbol{P}\epsilon R^{N_t \times K}$ and $\boldsymbol{\gamma}\epsilon R^{K \times 1}_+$}
\STATE {\textbf{repeat}}
\STATE \hspace{0.5cm}{calculate gradients in (\ref{grad-p}) and (\ref{grad-gamma})}
\STATE \hspace{0.5cm}{update $\boldsymbol{P}$ and $\boldsymbol{\gamma}$ using (\ref{iter-p}) and (\ref{iter-gamma})}
\STATE \hspace{0.5cm}{project $\boldsymbol{P}$ on intersection of manifolds as (\ref{proj_G1}-\ref{inter_sec})}
\STATE  {\textbf{until }convergence}
\STATE{\textbf{Output:} $\boldsymbol{P}$}
\end{algorithmic}
\end{algorithm}

\section{Simulation, Results, and Discussion}
To investigate the performance of the proposed algorithm, we consider SINR criteria and compare the results of the proposed method with ZF and MMSE precoding algorithms. We project the resultant precoding matrices of ZF and MMSE on the Manifolds (\ref{cons4}) and  (\ref{cons5}) to have a fair comparison with  the proposed algorithm. To have the impact of RIS in our proposed algorithm, we simulate another scenario without ORIS, namely "proposed without ORIS". In this scenario, an all-zero matrix is considered as the alignment matrix.
The room dimensions are considered $4$, $4$, and $3$ meters for width, length, and height, respectively. The LED array is mounted in the center of the room ceiling with $N_t=25$ LEDs arranged in a rectangular form. Also, the ORIS panel with various values of $M\epsilon\{24, 40, 64\}$ is placed at just the wall center. The users' locations are chosen randomly with a uniform distribution on the work plane and we calculate the SINRs by averaging over 1000 iterations. The other simulation parameters are summarized in table \ref{tab: parameters}.

\begin{table}
    \centering
\caption{Simulation parameters}
\label{tab: parameters}
    \begin{tabular}{|c|c|c|} \hline 
         \textbf{symbol}&  \textbf{description}& \textbf{value}\\ \hline 
         $A_p$&  PD area& $1 \text{ cm}^2$\\ \hline 
         $m$&  Lambert’s mode& $1$\\ \hline 
         $\theta_R$&  PD semi-angle& $\frac{\pi}{3}$\\ \hline 
         $T$&  signal transmission coefficien& $1$\\ \hline 
         $G$&  concentrator gain& $5$\\ \hline 
         $\alpha$&  reflective index of ORIS elements& $0.9$\\ \hline 
         $\mu$&  step size& $5e-4$\\ \hline
    \end{tabular}

\end{table}

\begin{figure}[!t]
\centering
\includegraphics[width=0.75\linewidth]{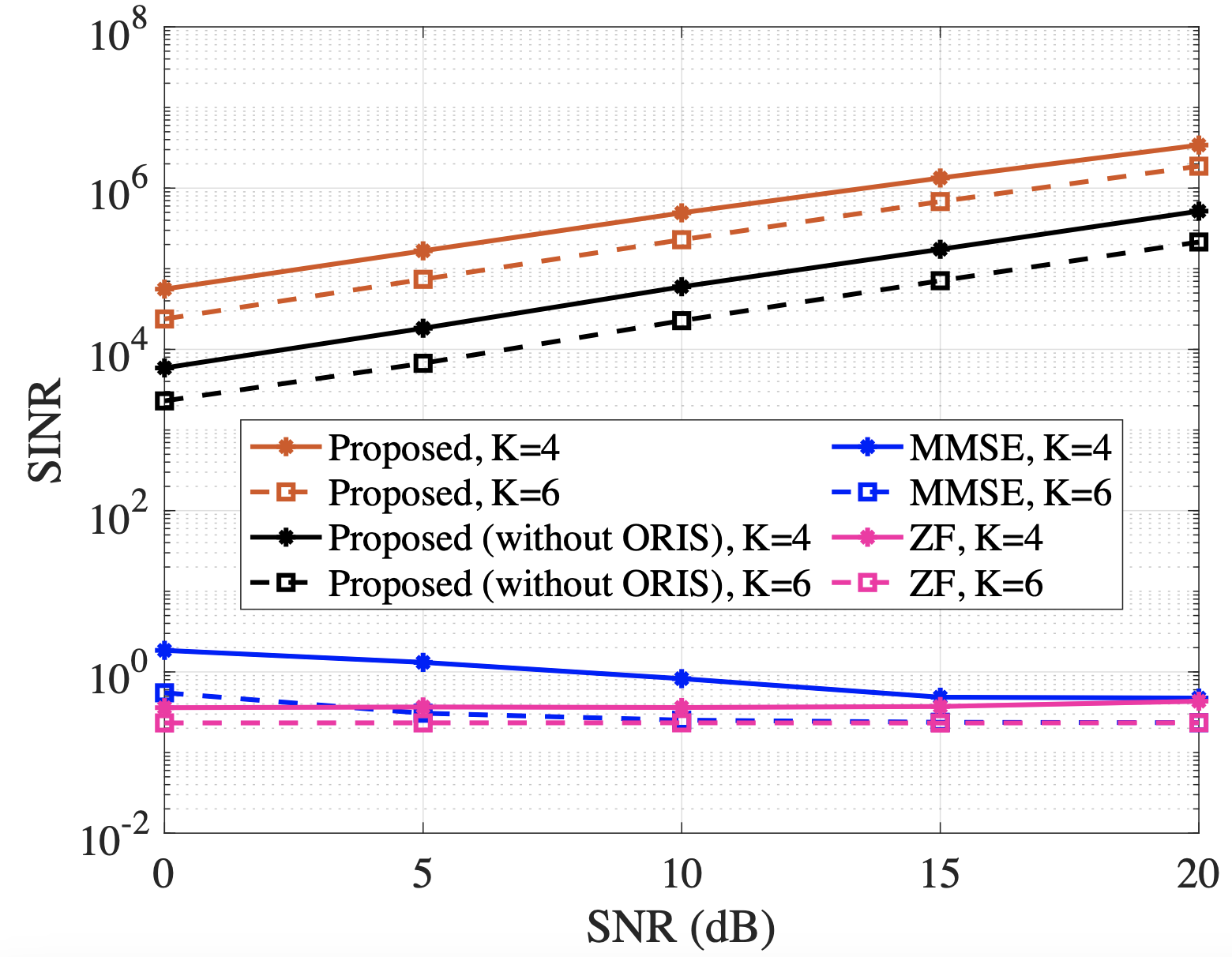}
\caption{The SINR curves for the number of users $K=4, 6$ for proposed, proposed without ORIS, MMSE, and ZF precoding algorithms.}
\label{sinr_k}
\end{figure}

\begin{figure}[!t]
\centering
\includegraphics[width=0.75\linewidth]{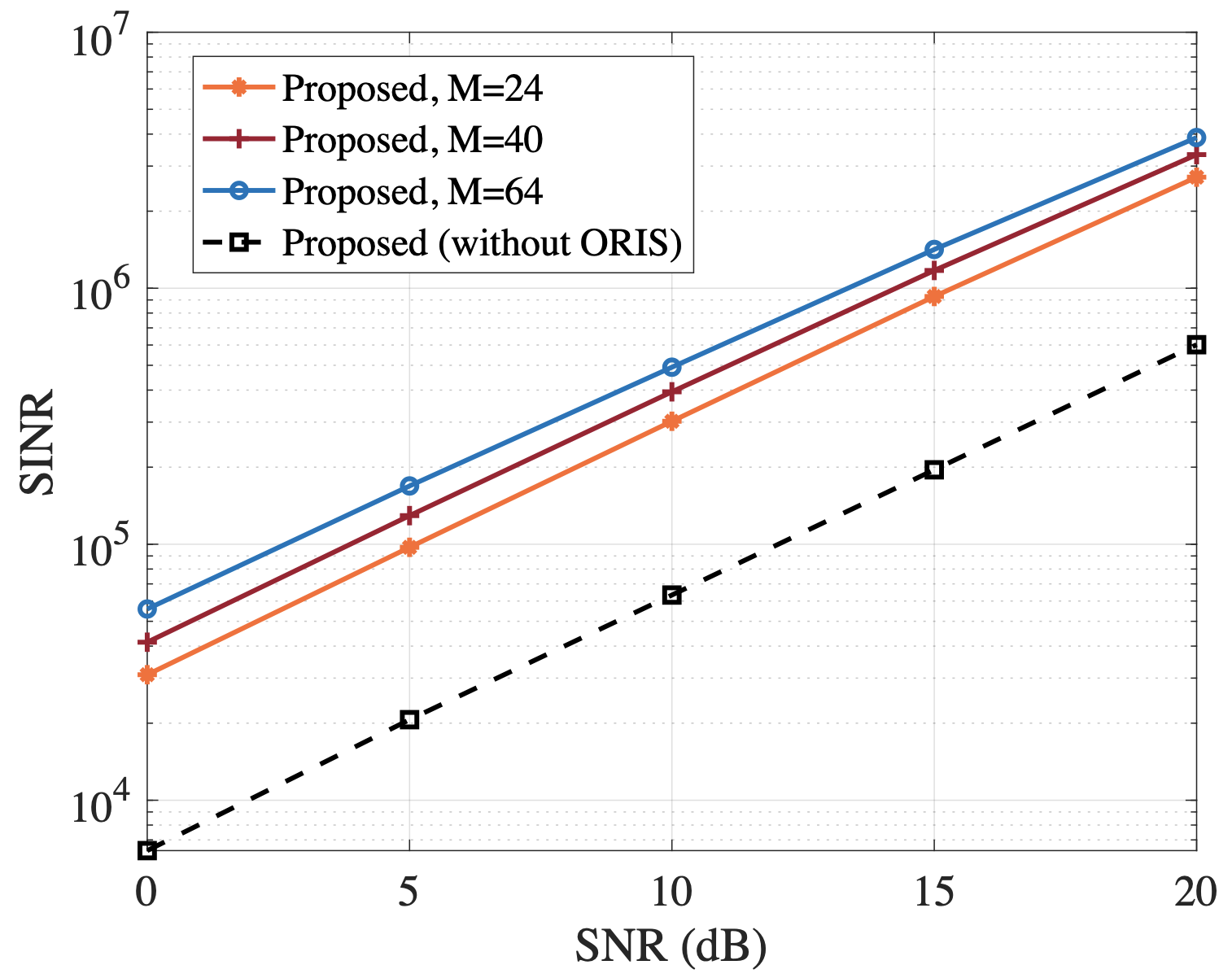}
\caption{The SINR curves under the number of ORIS elements $M=\{24, 40, 64\}$ for the proposed (and without ORIS) precoding algorithm.}
\label{sinr_n}
\end{figure}

Figure \ref{sinr_k} shows the SINR curves versus SNR for $K=4$ and 6. As seen, the proposed algorithm outperforms significantly the others. For instance, for $\text{SNR}=5\text{ dB}$, the $\text{SINR}$ values are $52.22, 42.62, 1.18, \text{and} -4.35 \text{ dBs}$ for proposed, proposed without ORIS, MMSE, and LS precoding algorithms under the presence of 4 users, which shows the $9.6 \text{ dB}$ improvement of proposed compared with proposed without ORIS as the impact of ORIS.
The poor performance of ZF and MMSE is because these algorithms have not been designed to satisfy the constraints (\ref{cons4}) and (\ref{cons5}). The SINR curves for $K=6$ are placed below those corresponding to $K=4$ since the total interference is increased by increasing user numbers. Moreover, as we expected, the presence of ORIS leads to higher SINR. For further investigation of the ORIS impact on SINR, the simulation results under various numbers of ORIS elements for the proposed precoding algorithm besides without ORIS one are shown in fig. \ref{sinr_n}. As one can see, increasing $N$ leads to better SINR. The larger number of ORIS elements improves signal strength at the receiver.

\section{Conclusion}
In this paper, we present an optimization problem to design precoding and alignment matrices jointly for the MU-MIMO VLC system in the presence of ORIS. To solve this problem we employed an alternative optimization algorithm, in which the gradient ascent projected on manifolds is used to design the precoding matrix. The simulation results show the better performance of the proposed algorithm in contrast to ZF and MMSE precoding algorithms from the SINR point of view.

\bibliographystyle{IEEEtran}

\bibliography{manuscript}

\vfill
\end{document}